\documentclass[aps,prd,reprint,superscriptaddress]{revtex4-2}

\usepackage{bm}
\usepackage{amssymb}
\usepackage{epsfig}
\usepackage{slashed}
\usepackage{color}
\usepackage{mathtools}
\usepackage{cases}
\usepackage{booktabs}
\usepackage{amsmath}
\usepackage{url}
\usepackage{graphicx}
\usepackage[
colorlinks=true,
filecolor=black,
anchorcolor=blue,
linkcolor=blue,
citecolor=cyan, 
urlcolor=cyan,
linktocpage=true,
plainpages=false,
breaklinks=true,
pdfstartview=FitH
]{hyperref}

\def\l{\left(}

\def\dd{\mathrm{d}}

\def\dd{\mathrm{d}}

\def\n{\hat{n}}

\def\T{\text{T}}
 
\def\E{\widetilde{E}}
\def\B{\widetilde{B}}

\def\l{\ell}

\newcommand{\bea}{\begin{eqnarray}}
\newcommand{\eea}{\end{eqnarray}}
\newcommand{\beq}{\begin{eqnarray}}
\newcommand{\eeq}{\end{eqnarray}}

\begin{document}





\title{Parity-odd CMB Correlations induced by a Chiral Gravitational Wave Background}

\author{Baoyu Xu}
\email[]{xuby@bao.ac.cn}
\affiliation{National Astronomical Observatories, Chinese Academy of Sciences, Beijing 100101, China}
\affiliation{School of Astronomy and Space Science, University of Chinese Academy of Sciences, Beijing 100049, China}

\author{Yun-Long Zhang}
\email[]{zhangyunlong@nao.cas.cn}
\affiliation{National Astronomical Observatories, Chinese Academy of Sciences, Beijing 100101, China}
\affiliation{School of Fundamental Physics and Mathematical Sciences, Hangzhou Institute for Advanced Study, University of Chinese Academy of Sciences, Hangzhou 310024, China}

\begin{abstract}
We investigate weak lensing effects of a chiral gravitational wave background (GWB) as a new mechanism for generating parity-odd correlations in the cosmic microwave background (CMB). Accounting for both photon deflection and polarization rotation, we derive the resulting parity-odd EB and TB angular power spectra. For representative GWB energy density spectra, we find parameter choices that provide good fits to existing measurements. This mechanism provides a potential observational probe of chiral GWB, which can arise in a wide range of theoretical scenarios.
\end{abstract}



\maketitle

\section{Introduction}

The parity-odd EB and TB correlations in the CMB provide sensitive probes of parity violation in cosmology. Current measurements leave room for nonzero values of these correlations, motivating investigations of their possible physical origins~\cite{Diego-Palazuelos:2022dsq,Diego-Palazuelos:2025dmh,Eskilt:2026imm}. One possible mechanism is cosmic birefringence, in which photon couplings to new particles, such as axions~\cite{Eskilt:2023nxm,Yin:2025fmj,Yin:2026gss} or dark photons~\cite{Lee:2023xzt}, rotate the linear polarization of CMB photons and generate parity-odd correlations. 
Moreover, chiral gravitational waves (GWs) can also induce EB and TB correlations via the radiative transfer mechanism~\cite{Lue:1998mq,Saito:2007kt,Hu:1997hp,Zaldarriaga:1996xe,gluscevic2010testing}. 
In addition to their direct contributions to CMB anisotropies, GWs can affect the propagation of CMB photons through weak lensing, inducing both photon deflection and polarization rotation and thereby generating additional B-mode polarization~\cite{Li:2006si,Dai:2013nda,Shaikh:2016hrf}. 
Building on these effects, we propose weak lensing by a chiral GWB as an alternative mechanism for generating CMB EB and TB correlations.  Accounting for both deflection and polarization rotation, we numerically calculate the induced angular power spectra. We find that, in the ultra-low frequency regime, our model can provide a good fit to existing measurements.
This mechanism offers a potential observational probe of chiral GWs, which can arise in a wide range of theoretical scenarios,  including inflation~\cite{Alexander:2004us,Li:2021wij,Li:2020xjt,Cai:2021uup,Takahashi:2009wc,Cai:2022lec,Ragavendra:2026fgs} and phase transitions~\cite{RoperPol:2019wvy,Ellis:2020uid,Gubler:2022zmf,Gorji:2026tln}. While these sources can produce signals across a broad range of frequencies, the lensing-induced correlations considered here offer a means of probing their parameter space in the ultra-low frequency regime.

In this paper, we first review the general formalism of weak lensing and extend it to the chiral GWB, deriving the lensing-induced corrections to the CMB angular power spectra, including the parity-odd EB and TB correlations. We then numerically evaluate these spectra for different GW energy density spectra and compare the predictions with existing observations. We find that chiral GWs with comoving wavenumbers in the range $k\sim 10^{-5}–10^{-2}\mathrm{Mpc}^{-1}$ can provide a good fit to the measured EB and TB spectra. Meanwhile, the associated B-mode power, which may exceed observational bounds for parameters favored by these fits, places an additional constraint on the amplitude of this contribution. This mechanism may therefore account for part of the observed EB signal, with its allowed contribution determined jointly by the EB, TB and B-mode measurements.

The structure of this paper is as follows. In section~\ref{Weak Lensing Effect}, we review the general formalism of weak lensing.  In section~\ref{EB and TB correlation}, we derive the expressions for the EB and TB angular power spectra in the CMB induced by a chiral GWB.  In section~\ref{correlation curves}, we numerically evaluate these spectra for different chiral GW energy density spectra and compare the results with existing observational data.  We place the summary and discussion at the end. Throughout this paper, we choose the metric signature $(-,+,+,+)$, and use natural units with $c=\hbar=1$.

\section{Weak Lensing Effect}\label{Weak Lensing Effect}
The propagation of CMB photons can be affected by gravitational lensing through two mechanisms. The first is the deflection of the photon propagation direction, which arises from the metric perturbations along the photon trajectory. It can be formally written as~\cite{Schmidt:2012ne}
\begin{align}
\begin{split}
    \delta& n^{i}(\n) = \Pi^{i}_{j}\big\{\frac{1}{2}\n^{k}(h_{o})^{j}_{k}-\beta^{j}_{o}
    -\int^{\chi_{s}}_{0} \dd\chi\big[(1-\frac{\chi}{\chi_{s}})
    \\
    &\times\big(\partial^{j}\alpha+\n^{k}\partial_{j}\beta_{k}-\frac{1}{2}\n^{k}\n^{l}\partial^{j}h_{kl}\big)-\frac{\beta^{j}-\n^{k}h^{j}_{k}}{\chi_{s}}\big]\big\}.
\end{split}
\end{align}
Here, $\Pi_{j}^{i}=\delta^{i}_{j}-\n^{i}\n_{j}$ is the projection operator, $\chi$ is the comoving radial distance from the observer along the line of sight, and $\alpha$ and $\beta_{i}$ represent scalar and vector perturbations, respectively. $h_{ij}$ denotes tensor perturbations, which reflect the influence of the GWB. From the above equation, scalar, vector, and tensor perturbations all induce directional perturbations in the propagation of CMB photons. However, since this paper focuses on the effects of chiral GWs on CMB photons, we set 
$\alpha=\beta_{i}=0$.

The other mechanism is polarization rotation, which arises from metric perturbations to the polarization direction of CMB photons~\cite{Dai:2013nda,Pitrou:2008hy,Liang:2025vji,Inomata:2025pnk}. It can be written as
\begin{align}
    \psi(\n) = \frac{1}{2} \epsilon^{ijk} \n_{k} \int^{\chi_{s}}_{0} \dd \chi \left(\partial_{i}\beta_{j}-n^{l}\partial_{i}h_{jl}\right).
\end{align}
Here, $\epsilon^{ijk}$ is the antisymmetric Levi-Civita tensor in three dimensional space. Only the transverse parts of vector and tensor perturbations can induce this mechanism. Similarly, in this work, we consider only the polarization rotation induced by tensor perturbations. 

These two mechanisms reveal how metric perturbations affect photon propagation in two ways. Not only by changing the propagation direction, but also by rotating the polarization direction that is perpendicular to it. By convention, the deflection $\delta n^{i}$ is decomposed into gradient and curl components in terms of the potentials $\phi(\n)$ and $\Omega(\n)$ as $\delta n^{i} = \nabla^{i}\phi-\epsilon^{ijk}\n_{j}\nabla_{k}\Omega$. For polarization rotation $\psi(\n)$, it can also induce a rotation of the CMB image as an effective curl potential. It has the following quantitative relationship with the curl potential
$2\psi=-\nabla^{2}\Omega$~\cite{Dodelson:2003bv,Dai:2013nda}. This relation reflects the common influence of metric perturbations on the deflection of the light rays and polarization transport. Although $\Omega$ and $\psi$ are numerically equivalent in this framework, they describe distinct physical effects. Therefore, their contributions should consequently be treated jointly, including their cross-correlations.

\section{Parity-odd CMB correlation}
\label{EB and TB correlation}

Photon deflection and polarization rotation modify the observed CMB temperature and polarization fields.
Firstly, regarding the polarization measurements of CMB photons, the gradient and curl potentials induced by directional deflection affect the Stokes parameters $Q(\n)$ and $U(\n)$, while polarization rotation influences the mixing between them. Their combined effect can be written as $Q(\n)\pm iU(\n) \Rightarrow e^{\mp2 i \psi(\n)}\left[Q(\n+\delta n)\pm iU(\n+\delta n)\right]$. These effects modify the CMB E- and B-mode polarization and their correlations. In the presence of a chiral GWB, they can generate a nonzero parity-odd EB correlation. For CMB temperature, photon deflection remaps the temperature anisotropies as $\T(\n)\Rightarrow\T(\n+\delta n)$, while polarization rotation leaves the temperature field unchanged. Together with the modified B-mode polarization, this remapping can give rise to a nonzero parity-odd TB correlation in the presence of a chiral GWB. 

Both the photon deflection and polarization rotation modify the CMB E- and B-mode polarization, leading to corrections to the observed angular power spectra. We assume that the unlensed EB and TB spectra vanish, $C^{EB}_{\l'}=C^{\T B}_{\l'}=0$, and neglect any sources of parity-odd correlations other than lensing by chiral GWs.  Retaining terms up to second order in the lensing perturbations, we obtain the induced EB angular power spectrum as
\begin{align}\label{C_EB}
    C^{EB}_{\l} = -&\frac{1}{2\l+1}\sum_{\l' L}(-1)^{\l+\l'+L}\bigg\{C^{\Omega\phi}_{L} G^{\Omega}_{\l\l'L}G^{\phi}_{\l\l'L}\nonumber
    \\
    &-2C^{\phi\psi}_{L} H^{\psi}_{\l\l'L}G^{\phi}_{\l\l'L}\bigg\}
    \left[C^{EE}_{\l'}-C^{BB}_{\l'}\right].
\end{align}
Under the same assumptions, the TB angular power spectrum induced by chiral GW lensing is given by
\begin{align}\label{C_TB}
    &C^{\T B}_{\l} = -\frac{1}{2\l+1} \sum_{\l'L}\bigg\{C^{\phi\Omega}_{L}G^{\phi}_{\l\l'L}K_{\l\l'L}C^{\T E}_{\l'}P^{-}_{\l\l'L}\nonumber
    \\
    &-\left[C^{\phi\Omega}_{L}G^{\Omega}_{\l\l'L}K_{\l\l'L}
    -2C^{\phi\psi}_{L}H^{\psi}_{\l\l'L}K_{\l\l'L}\right]C^{\T E}_{\l'}P^{+}_{\l\l'L}\bigg\}.
\end{align}
Here, we have defined $P^{\pm}_{\l\l'L}=[1\pm(-1)^{\l+\l'+L}]/2$ as parity projectors that select even and odd values of $[\l+\l'+L]$, respectively. 
And the four coupling kernels $G^{\phi}_{\l\l'L}$,  $G^{\Omega}_{\l\l'L}$, $H^{\psi}_{\l\l'L}$ and $K_{\l\l'L}$ are given by
\begin{align}
\begin{split}
   H^{\psi}_{\l\l'L} &= \sqrt{\frac{\Pi_{\l\l'L}}{4\pi}}\begin{pmatrix}
\l&L&\l'\\
2&0&-2\end{pmatrix}
\\
G^{\phi / \Omega}_{\l\l'L} &= \sqrt{\frac{L(L+1)\Pi_{\l\l'L}}{16\pi}}\big[\sqrt{(\l'+2)(\l'-1)}
    \begin{pmatrix}
    \l&L&\l'\\
    2&-1&-1\end{pmatrix}
\\
&\pm \sqrt{(\l'-2)(\l'+3)}
\begin{pmatrix}
\l&L&\l'\\
2&1&-3\end{pmatrix}\big],
\\
K_{\l\l'L}&=-\sqrt{\frac{L(L+1)\l'(\l'+1)\Pi_{\l\l'L}}{4\pi}}\begin{pmatrix}
\l&L&\l'\\
0&-1&1\end{pmatrix},
\end{split}
\end{align}
where matrices denote the Wigner-3j symbols, with the factor $\Pi_{\l\l'L}=(2\l+1)(2\l'+1)(2L+1)$. Moreover, $C^{\phi\Omega}_{L}$ and  $C^{\phi\psi}_{L}$ are parity-odd correlations, which are related to the energy-density spectrum and chirality of the chiral GWB.  In particular, $C^{\phi\Omega}_{L}$ can be written as~\cite{Inomata:2024ald,book2012odd} 
\begin{align}\label{C_phi_omega}
    C^{\phi\Omega}_{L} = \int \frac{k^{2}\dd k}{2\pi^{2}}\left[P_{\text{L}}(k)-P_{\text{R}}(k)\right]F^{\phi}_{L}F^{\Omega}_{L},
\end{align}
where $F^{\Omega}_{L}(k) = N_{L} \int^{k\eta_0}_{k\eta_{r}} T(\omega)j_{L}(x)/x^{2}\dd \omega $ and $F_{L}^{\phi}(k) = -N_{L}\int^{k\eta_0}_{k\eta_{r}}\dd\omega [T(\omega)/k(\eta_{0}-\eta_{r})]
\big\{x[-(\partial/\partial x)+(y/2)(1+\partial^{2}/\partial x^{2})]-3+2y(\partial/\partial x)\big\}(j_{L}(x)/x^{2})$. $x=k\eta_{0}-\omega$ and $y=\omega-k\eta_{r}$, with $\omega=k\eta$ is the dimensionless wavenumber. Here, $\eta_0$ and $\eta_r$ denote the conformal times today and at recombination, respectively, which means $\chi_{s}=\eta_{0}-\eta_{r}$. Moreover, the normalization factor is $N_{L}=[1/L(L+1)]\sqrt{2\pi[(L+2)!/(L-2)!]}$ and we adopt the GW transfer function $T(\omega)=3j_{1}(\omega)/\omega$.
Using $2\psi=-\nabla^{2}\Omega$, we obtain $2\psi_{LM}=L(L+1)\Omega_{LM}$ in harmonic space. 
The corresponding cross-power spectra therefore satisfy $2C^{\phi\psi}_{L} = L(L+1)C^{\phi\Omega}_{L}$. Finally, defining the dimensionless chiral power spectrum $\mathcal{P}_{V}(k)=(k^{3}/2\pi^{2})\left[P_{\text{L}}(k)-P_{\text{R}}(k)\right]$  to characterize the helicity asymmetry of the chiral GWB, we obtain $2C^{\phi\psi}_{L}=L(L+1)\int  \mathcal{P}_{V}(k)F^{\phi}_{L}(k)F^{\Omega}_{L}(k)\dd \text{ln}k$. 

As shown in Eq.~\eqref{C_EB}, Eq.~\eqref{C_TB}, and Eq.~\eqref{C_phi_omega}, a chiral GWB generates the parity-odd cross correlations $C_{L}^{\phi\Omega}$ and $C_{L}^{\phi\psi}$ between the lensing fields, which in turn induce CMB EB and TB correlations through photon deflection and polarization rotation. These expressions therefore allow us to calculate the induced angular power spectra for a given chiral GW power spectrum, providing a connection between GW chirality and parity-odd CMB observables.

\section{EB and TB from Chiral GWBs}
\label{correlation curves}

We consider the CMB EB and TB angular power spectra induced by two representative forms of chiral GW energy spectra from cosmological sources: the flat spectrum and the single peaked spectrum.  Since many cosmological GW spectra exhibit power-law behavior, we parameterize the dimensionless chiral power spectrum $\mathcal{P}_{V}(k)$ using a broken power-law (BPL) model, which accommodates both spectral shapes through suitable choices of its parameters. The expression of the BPL model is given by~\cite{Kuroyanagi:2018csn,Su:2025nkl}
\begin{align}
    \mathcal{P}^{\text{BPL}}_{V}=\sigma_{\text{L,R}}\mathcal{P}_{c}\left(\frac{k}{k_{c}}\right)^{\alpha_{1}}\left[1+0.75\left(\frac{k}{k_{c}}\right)^{\Delta}\right]^{\frac{\alpha_{2}-\alpha_{1}}{\Delta}}.
\end{align}
Here, $\mathcal{P}c$ is the amplitude parameter, $k_c$ is the characteristic comoving wavenumber, and $\sigma_{\mathrm{L,R}}=\pm1$ specifies left- and right-handed chirality, respectively. For $\Delta>0$, the parameters $\alpha_1$ and $\alpha_2$ specify the asymptotic slopes at low and high wavenumbers, respectively, while $\Delta$ controls the sharpness of the transition. When $\alpha_{1}=\alpha_{2}=\alpha$, the BPL model reduces to $\mathcal{P}^{\text{BPL}}_{V}=\sigma_{\text{L,R}}\mathcal{P}{c}\left(k/k_{c}\right)^{\alpha}$, which describes  a flat spectrum. Furthermore, by choosing $\alpha_{1}>0$ and $\alpha_{2}<0$, the BPL model can reproduce a single peaked spectrum.  For each spectral model, we use the same parameter set to calculate the CMB EB and TB angular power spectra and compare them with their respective observational data.

\begin{figure}[h]
\includegraphics[scale=0.43]{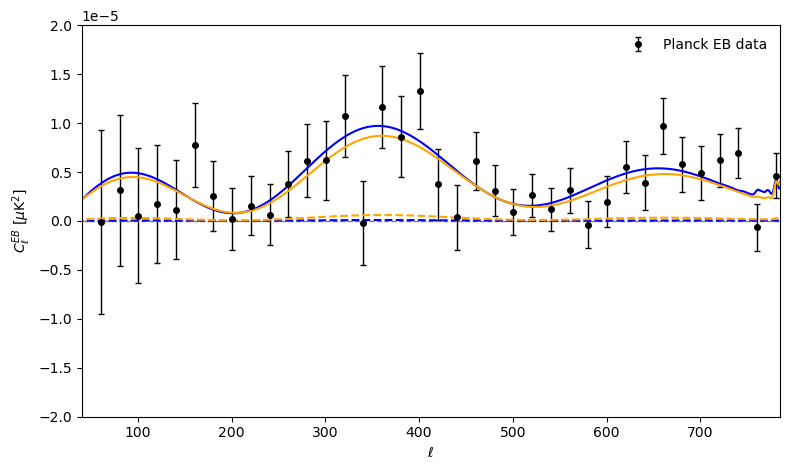}
\caption{\label{C_EB_fig} CMB EB angular power spectra induced by chiral GW lensing for the flat spectrum model (blue curve) and the single-peak model (orange curve). The data are the observed EB power spectrum from the High Frequency Instrument of Planck~\cite{Eskilt:2023nxm,Planck:2020olo}. The solid curves show representative fits, yielding $\chi^2\simeq25$ and $\chi^2\simeq24$ for the flat spectrum and single-peak models, respectively, over 36 bins. The dashed curves show the corresponding predictions after reducing the GW amplitudes to satisfy the BB constraints.} 
\end{figure}

\begin{figure}[h]
\includegraphics[scale=0.43]{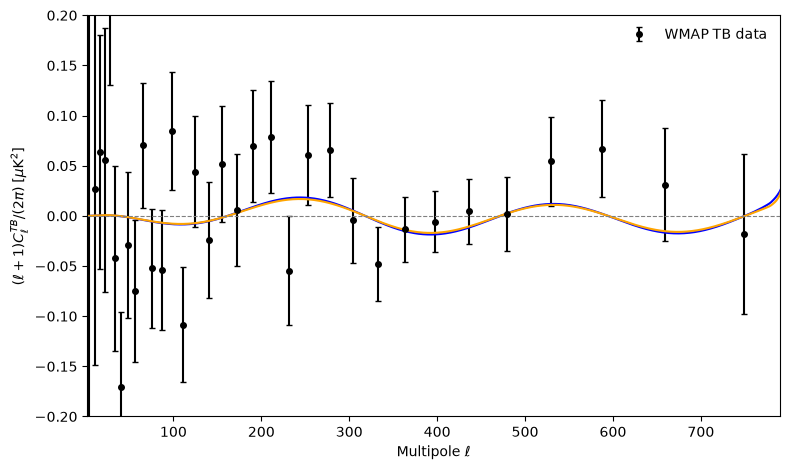}
\caption{CMB TB angular power spectra induced by chiral GW lensing for the flat spectrum (blue) and single-peaked (orange) models, using the same parameter values as in the EB fits. 
The data used is the WMAP 9-year TB data~\cite{Bennett2013WMAP9}. The predicted nonzero TB spectra provide a better fit to the data than the null hypothesis of vanishing TB correlations, with both models yielding  $\chi^2\simeq34$.
\label{C_TB_fig}}
\end{figure}

For the flat spectrum, with $\alpha_{1}=\alpha_2=0$ , fitting the EB data alone yields a best fit amplitude of $\mathcal{P}_c=0.08$ with $\sigma_{R}=-1$.  Although the input spectrum is scale invariant, the lensing kernels select a finite range of scales, with GWs at $k\sim10^{-5}$--$10^{-2}\mathrm{Mpc}^{-1}$ providing the dominant contribution to the induced EB signal.  The relevant integrals converge at both low and high wavenumbers, and the resulting EB angular power spectrum is shown in Fig.~\ref{C_EB_fig}.  For the single-peak model, several parameter combinations provide comparably good fits to the EB data. As a representative example, we adopt  $\alpha_{1}=10$, $\alpha_2=-2$, $k_c=10^{-3}\text{Mpc}^{-1}$, $\mathcal{P}_{c}=0.2$ with $\sigma_{R}=-1$ and $\Delta=10$. The corresponding EB angular power spectrum is also shown in Fig.~\ref{C_EB_fig}.
Within the parameter space considered, our model provides a good fit to the observed EB spectrum. For sufficiently long-wavelength GW modes that remain well outside the horizon until recombination, the amplitudes are approximately constant, suppressing the tensor source terms responsible for generating CMB anisotropies at that epoch~\cite{Caprini:2018mtu,Kite:2021yoe}. These modes begin to evolve appreciably upon horizon entry. Here, we focus on the weak lensing effects accumulated during subsequent CMB photon propagation, adopting the matter-era transfer function to describe the GW evolution.
Using the same parameter values, we calculate the corresponding TB angular power spectra, shown in Fig.~\ref{C_TB_fig}. These predictions also provide a better fit to the TB data than the null hypothesis of vanishing TB correlations.

The representative parameter choices above provide good fits to the EB data, but may also lead to BB power that is in tension with existing observational bounds~\cite{Shaikh:2016hrf,POLARBEAR:2022dxa,SPT:2019nip}. Although curl deflection and polarization rotation partially cancel in their contributions to the BB spectrum~\cite{Dai:2013nda}, the residual power suggests that a reduction in the GW amplitude may be needed. Keeping all other parameters fixed, we reduce $\mathcal{P}_c$ to $1.6\times10^{-3}$ for the flat spectrum model and $0.014$ for the single-peak model, bringing the predictions into better agreement with the BB constraints. The corresponding EB spectra are shown as dashed curves in Fig.~\ref{C_EB_fig}. With these reduced amplitudes, the mechanism can still contribute to the EB signal, although it may not account for its full amplitude. Weak lensing by a chiral GWB therefore remains a possible source of a fraction of the observed EB correlation.

\section{Conclusion}

In this work, we have investigated the weak lensing effects of a chiral gravitational wave background on the CMB, including both photon deflection and polarization rotation. We have derived the resulting corrections to the CMB angular power spectra and shown that these propagation effects can generate parity-odd EB and TB correlations. Our results identify weak lensing by chiral GWs as a possible mechanism for producing such correlations, in addition to the direct contributions of tensor perturbations to CMB anisotropies.

By numerically evaluating the induced spectra for different GW energy density spectra, we find that chiral GWs with comoving wavenumbers in the range $k\sim 10^{-5}$-$10^{-2}\mathrm{Mpc}^{-1}$ can provide a good fit to existing EB and TB measurements. At the same time, the associated B-mode power may exceed observational constraints for the parameters required by these fits. The agreement with EB and TB measurements must therefore be assessed together with the predicted BB spectrum. This mechanism may still contribute to a fraction of the observed EB signal, although the allowed contribution remains to be quantified through a joint analysis.

These results suggest that parity-odd CMB correlations offer a potential probe of chiral GWB in the ultra-low-frequency regime. Since chiral GWs can arise in a variety of early universe scenarios, the propagation effects studied here provide an additional connection between their properties and CMB observables. Their application to specific production models requires a consistent treatment of the predicted GW spectrum and the associated CMB signals. In addition, chiral GWs may contribute to CMB temperature anisotropies through the tensor Sachs–Wolfe effect~\cite{Sachs:1967er,Nishizawa:2014vga}, potentially giving rise to an additional TB contribution through correlations with the lensing-induced B modes. We leave a detailed investigation of this contribution to future work.

\vspace{10pt}

\begin{acknowledgments}
This work is supported by the
National Natural Science Foundation of China (No.12375059), the Project of National Astronomical Observatories, Chinese Academy of Sciences (No.E4TG6601).
\end{acknowledgments}


\onecolumngrid
\appendix

\phantomsection
\addcontentsline{toc}{section}{\protect\numberline{}Appendix}

\let\savedaddcontentsline\addcontentsline
\renewcommand{\addcontentsline}[3]{}

\section{CMB Polarization in Chiral Gravitational Waves} 
The gradient potential, curl potential, and polarization rotation caused by gravitational wave lensing collectively act on CMB photons to remap the CMB polarization and mix its E and B modes, which can be expressed as ${}_{\pm2}P(\n)=Q(\n)\pm iU(\n) \Longrightarrow e^{\mp2 i \psi(\n)}\left[Q(\n+\delta n)\pm iU(\n+\delta n)\right]  ={}_{\pm2}\widetilde{P}(\n)$. Furthermore, it can naturally be decomposed into E and B modes via spherical harmonics as follows,
\begin{align}
    {}_{\pm2}P(\n) = \sum_{\l m}\left(E_{\l m}\pm i B_{\l m}\right){}_{\pm2}Y_{\l m}(\n) \Longrightarrow
    {}_{\pm2}\widetilde{P}(\n) = \sum_{\l m}\left(\widetilde{E}_{\l m}\pm i \widetilde{B}_{\l m}\right){}_{\pm2}Y_{\l m}(\n).
\end{align}
Expanding ${}_{\pm2}\widetilde{P}(\n)$ perturbatively and considering terms up to second order, we can obtain ${}_{\pm2}\widetilde{P}(\n) = {}_{\pm2}\widetilde{P}^{(0)}(\n) + {}_{\pm2}\widetilde{P}^{(1)}(\n) + {}_{\pm2} \widetilde{P}^{(2)}(\n)+\cdots.$ Here, we  defined ${}_{\pm2}\widetilde{P}^{(n)}(\n) = \sum_{\l m}\left(\E^{(n)}_{\l m}\pm i \B^{(n)}_{\l m}\right){}_{\pm2}Y_{\l m}(\n)$ and it is straightforward to obtain the relation between the ${}_{\pm2}\widetilde{P}^{(n)}(\n)$ and $\E,\B$ that
$\left(\widetilde{E}^{(n)}_{\l m}\pm i \widetilde{B}^{(n)}_{\l m}\right)= \int \dd \n \, {}_{\pm2}Y^{*}_{\l m}(\n){}_{\pm2}\widetilde{P}^{(n)}(\n)$ with
\begin{align}
    \widetilde{E}^{(n)}_{\l m} &= \frac{1}{2}\int \dd \n  \left[{}_{+2}Y^{*}_{\l m}(\n){}_{+2}\widetilde{P}^{(n)} + {}_{-2}Y^{*}_{\l m}(\n){}_{-2}\widetilde{P}^{(n)}\right],
    \\
    i\widetilde{B}^{(n)}_{\l m} &= \frac{1}{2}\int \dd \n  \left[{}_{+2}Y^{*}_{\l m}(\n){}_{+2}\widetilde{P}^{(n)} - {}_{-2}Y^{*}_{\l m}(\n){}_{-2}\widetilde{P}^{(n)}\right].
\end{align}
Moreover, expanding ${}_{\pm2}\widetilde{P}$, we have
\begin{align}
    {}_{\pm2}\widetilde{P}(\n)&= (1\mp2 i \psi -2\psi^{2}+\cdots)(_{\pm 2}P(\n)+\delta n^{i}\nabla_{i} \ _{\pm 2}P(\n)+\frac{1}{2}\delta n^{i}\delta n^{j}\nabla_{i}\nabla_{j} \ _{\pm 2}P(\n)+\cdots) \nonumber
    \\
    & = _{\pm 2}P(\n) +  \delta n^{i}\nabla_{i} \ _{\pm 2}P(\n)\mp2 i \psi _{\pm 2}P(\n) + \frac{1}{2}\delta n^{i}\delta n^{j}\nabla_{i}\nabla_{j}\ _{\pm 2}P(\n) \mp2 i \psi \delta n^{i}\nabla_{i} \ _{\pm 2}P(\n) -2\psi^{2} \ _{\pm 2}P(\n)+\cdots,
\end{align}
which we can extract the components from zeroth to second order as ${}_{\pm2}\widetilde{P}^{(0)}(\n)=\ _{\pm 2}P(\n)$, ${}_{\pm2}\widetilde{P}^{(1)}(\n) = \delta n^{i}\nabla_{i} \ _{\pm 2}P(\n)\mp2 i \psi _{\pm 2}P(\n) $ and ${}_{\pm2}\widetilde{P}^{(2)}(\n) =\frac{1}{2}\delta n^{i}\delta n^{j}\nabla_{i}\nabla_{j}\ _{\pm 2}P(\n) \mp2 i \psi \delta n^{i}\nabla_{i} \ _{\pm 2}P(\n) -2\psi^{2} \ _{\pm 2}P(\n)$ with $\delta n^{i}=\nabla^{i} \phi - \epsilon^{ijk}n_{j}\nabla_{k}\Omega$. Therefore, substituting $\delta n^{i}$ into $_{\pm 2}\widetilde{P}^{(n)}$ , we obtain
\begin{align}
    {}_{\pm2}\widetilde{P}^{(0)}(\n) &=\ _{\pm 2}P(\n) = \sum_{\l' m'}\left(E_{\l' m'}\pm i B_{\l' m'}\right){}_{\pm2}Y_{\l' m'}(\n),
    \\
    {}_{\pm2}\widetilde{P}^{(1)}(\n) 
    &=\sum_{LM}\sum_{\l' m'}\phi_{LM}(E_{\l' m'}\pm iB_{\l' m'})\nabla^{i}Y_{LM}\nabla_i{}_{\pm2}Y_{\l' m'}\nonumber
\\ 
&\qquad-\sum_{LM}\sum_{\l' m'}\Omega_{LM}(E_{\l' m'}\pm iB_{\l' m'})\epsilon^{ijk}n_j\nabla_kY_{LM}\nabla_i{}_{\pm2}Y_{\l' m'}\nonumber
\\&\qquad\mp2i\sum_{LM}\sum_{\l' m'}\psi_{LM}(E_{\l' m'}\pm iB_{\l' m'})Y_{LM}\,{}_{\pm2}Y_{\l' m'},
\\
{}_{\pm2}\widetilde{P}^{(2)}(\n) &=-\frac{1}{2}\sum_{\l' m'}\left\{[\l'(\l'+1)-4]R+4S\right\} \left(E_{\l'm'}\pm i B_{\l'm'}\right){}_{\pm2}Y_{\l'm'}\,.\nonumber
\end{align}
Here we set $\langle\delta n^{i}\delta n_{i}\rangle=2R$, $\langle\psi^{2}\rangle=S$, and using the symmetry, we obtain $\langle\psi\delta n^{i}\rangle = 0$. Then, use $\left(\widetilde{E}_{\l m}\pm i \widetilde{B}_{\l m}\right)= \int \dd \n \left[{}_{\pm2}Y^{*}_{\l m}(\n){}_{\pm2}\widetilde{P}(\n)\right]$, we have
\begin{align}
    \E^{(0)}_{\l m}\pm i\B^{(0)}_{\l m} &= \sum_{\l' m'}\left(E_{\l' m'}\pm i B_{\l' m'}\right) \int \dd \n \, \left[{}_{\pm2}Y^{*}_{\l m}(\n){}{}_{\pm2}Y_{\l' m'}(\n)\right],\label{E0}
    \\
    \E^{(1)}_{\l m}\pm i\B^{(1)}_{\l m}&=
    \sum_{LM\l' m'}\phi_{LM}(E_{\l'm'}\pm iB_{\l'm'})\int \dd\n\left[{}_{\pm2}Y^*_{\l m}\nabla^{i}Y_{LM}\nabla_i{}_{\pm2}Y_{\l' m'}\right]\nonumber
    \\
    &\qquad-\sum_{LM\l' m'}\Omega_{LM}(E_{\l' m'}\pm iB_{\l' m'})\int \dd\n\left[{}_{\pm2}Y^*_{\l m}\epsilon^{ijk}n_j\nabla_kY_{LM}\nabla_i{}_{\pm2}Y_{\l' m'}\right]\nonumber
    \\
    &\qquad\mp2i\sum_{LM\l'm'}\psi_{LM}(E_{\l'm'}\pm iB_{\l'm'})\int \dd\n\left[{}_{\pm2}Y^*_{\l m}Y_{LM}{}_{\pm2}Y_{\l' m'}\right],
    \\
    \E^{(2)}_{\l m}\pm i\B^{(2)}_{\l m}&=-\frac{1}{2}\sum_{\l'm'}\left[(\l'(\l'+1)-4)R+4S\right]\left(E_{\l'm'}\pm i B_{\l'm'}\right)\int \dd \n \, \left[{}_{\pm2}Y^{*}_{\l m}(\n){}{}_{\pm2}Y_{\l'm'}(\n)\right].\label{E2}
\end{align}
Using relation $\int \dd \n \left[{}_{\pm2}Y^{*}_{\l m}(\n){}{}_{\pm2}Y_{\l'm'}(\n)\right]=\delta_{\l\l'}\delta_{mm'}$, we can conveniently calculate the zeroth- and second-order perturbed E and B modes, which satisfy simple relations with the unperturbed E and B modes, that $\E^{(0)}_{\l m}=E_{\l m}$, $\B^{(0)}_{\l m}=B_{\l m}$ and $\E^{(2)}_{\l m}=-\frac{1}{2}\left[(\l(\l+1)-4)R+4S\right]E_{\l m}$, $\B^{(2)}_{\l m}=-\frac{1}{2}\left[(\l(\l+1)-4)R+4S\right]B_{\l m}$. Subsequently, we define three additional integral kernels as
\begin{align}
    {}_{\pm2}I^{\psi}_{\l\l'LM}&=\int \dd\n \left[{}_{\pm2}Y^*_{\l m}Y_{LM}{}_{\pm2}Y_{\l'm'}\right],
\\
    {}_{\pm2}I^{\phi}_{\l\l'LM}&=\int \dd\n \left[{}_{\pm2}Y^*_{\l m}\nabla^{i}Y_{LM}\nabla_i{}_{\pm2}Y_{\l'm'}\right],
\\
    {}_{\pm2}I^{\Omega}_{\l\l'LM}&=\int \dd\n\left[{}_{\pm2}Y^*_{\l m}\epsilon^{ijk}n_j\nabla_kY_{LM}\nabla_i{}_{\pm2}Y_{\l' m'}\right].
\end{align}
Thus, the first-order perturbed E- and B-modes can be expressed as
\begin{align}
\E^{(1)}_{\l m}
=\frac{1}{2}&\sum_{LM\l'm'}\phi_{LM}\left[E_{\l'm'}({}_{+2}I^\phi+{}_{-2}I^\phi)+iB_{\l'm'}({}_{+2}I^\phi-{}_{-2}I^\phi)\right]\nonumber
\\
-\frac{1}{2}&\sum_{LM\l'm'}\Omega_{LM}\left[E_{\l'm'}({}_{+2}I^\Omega+{}_{-2}I^\Omega)+iB_{\l'm'}({}_{+2}I^\Omega-{}_{-2}I^\Omega)\right]\nonumber
\\
-i&\sum_{LM\l'm'}\psi_{LM}\left[E_{\l'm'}({}_{+2}I^\psi-{}_{-2}I^\psi)+iB_{\l'm'}({}_{+2}I^\psi+{}_{-2}I^\psi)\right],\label{E1}
\end{align}
\begin{align}
\B^{(1)}_{\l m}=\frac12&\sum_{LM\l'm'}\phi_{LM}\left[-iE_{\l'm'}({}_{+2}I^\phi-{}_{-2}I^\phi)+B_{\l'm'}({}_{+2}I^\phi+{}_{-2}I^\phi)\right]\nonumber
\\
-\frac12&\sum_{LM\l'm'}\Omega_{LM}\left[-iE_{\l'm'}({}_{+2}I^\Omega-{}_{-2}I^\Omega)+B_{\l'm'}({}_{+2}I^\Omega+{}_{-2}I^\Omega)\right]\nonumber
\\
-i&\sum_{LM\l'm'}\psi_{LM}\left[-iE_{\l'm'}({}_{+2}I^\psi+{}_{-2}I^\psi)+B_{\l'm'}({}_{+2}I^\psi-{}_{-2}I^\psi)\right].\label{B1}
\end{align}

\subsection{Integral kernels}
\label{Integral Kernels}
In this subsection, we calculate the three integral kernels ${}_{\pm2}I^{\psi}_{\l\l'LM}$, ${}_{\pm2}I^{\phi}_{\l\l'LM}$ and ${}_{\pm2}I^{\Omega}_{\l\l'LM}$. Before evaluating these integrals, we need to introduce several formulas. One is the Gaunt integral, we can use $\int d\Omega\;{}_{s_1}Y^*_{l_1m_1}\,{}_{s_2}Y_{l_2m_2}\,{}_{s_3}Y_{l_3m_3}=
{}(-1)^{m_1+s_1}
\sqrt{\frac{\Pi_{l_1l_2l_3}}{4\pi}}
\begin{pmatrix}l_1&l_2&l_3\\
-s_1&s_2&s_3\end{pmatrix}
\begin{pmatrix}l_1&l_2&l_3\\
-m_1&m_2&m_3\end{pmatrix}$ with $\Pi_{l_{1}l_{2}l_{3}}=(2l_{1}+1)(2l_{2}+1)(2l_{3}+1)$. The other is the Newman-Penrose formalism, we have $\nabla_a=\frac1{\sqrt2}\left(m_a\bar\eth+\bar m_a\eth\right)$ and $\epsilon^{ijk}n_j\nabla_k=-\frac{i}{\sqrt2}\left(m^i\bar\eth-\bar m^i\eth
\right)$ with $m^{a}m_{a}=0,\bar m_a\bar m^a=0,m^a \bar m_{a}=1$, where  $\eth$ and $\bar\eth$ are the spin-raising and spin-lowering operators, respectively, which satisfy $\eth\,{}_sY_{lm}=\sqrt{(l-s)(l+s+1)}\,{}_{s+1}Y_{lm}$,
$\bar\eth\,{}_sY_{lm}=-\sqrt{(l+s)(l-s+1)}\,{}_{s-1}Y_{lm}.$ Using these, we can conveniently compute the three integral kernels. 

First, for the integral kernel  ${}_{\pm2}I^{\psi}_{\l\l'LM}$, we only need to use Gaunt integral, where
\begin{align}
    {}_{\pm2}I^{\psi}_{\l\l'LM}&=\int \dd\n\left[{}_{\pm2}Y^*_{\l m}Y_{LM}{}_{\pm2}Y_{\l'm'}\right] = \mathcal{W}_{lLl'}\sqrt{\frac{\Pi_{ll'L}}{4\pi}}
\begin{pmatrix}
\l&L&\l'\\
\mp2&0&\pm2\end{pmatrix}.
\end{align}
Here, we defined $\mathcal{W}_{\l L \l'}=(-1)^{m}\begin{pmatrix}
\l&L&\l'\\
-m&M&m'\end{pmatrix}$. Then, we have integral kernel  ${}_{\pm2}I^{\phi}_{\l\l'LM}$ 
\begin{align}
    {}_{\pm2}I^{\phi}_{\l\l'LM}&=\int \dd\n\left[{}_{\pm2}Y^*_{lm}\nabla^{i}Y_{LM}\nabla_i{}_{\pm2}Y_{l'm'}\right]=\frac{1}{2}\int \dd\n\,{}_{\pm2}Y^*_{\l m}\left(\bar\eth Y_{LM}\eth{}_{\pm2}Y_{\l' m'}+\eth Y_{LM}\bar\eth{}_{\pm2}Y_{\l' m'}\right),
\\
    {}_{\pm2}I^{\Omega}_{\l\l'LM}&=\int \dd\n\left[{}_{\pm2}Y^*_{\l m}\epsilon^{ijk}n_j\nabla_kY_{LM}\nabla_i{}_{\pm2}Y_{\l'm'}\right]=-\frac{i}{2}\int \dd\n \,{}_{\pm2}Y^*_{\l m}\left(\bar\eth Y_{LM}\eth{}_{\pm2}Y_{\l'm'}-\eth Y_{LM}\bar\eth{}_{\pm2}Y_{\l' m'}\right).
\end{align}
Upon expanding the equation, we arrive at
\begin{align}
    {}_{+2}I^{\phi}_{\l\l'LM}&=-(-1)^{\l+L+\l'}\mathcal{W}_{\l L\l'}\sqrt{\frac{\Pi_{\l\l'L}L(L+1)}{8\pi}}
\big[\sqrt{\frac{(\l'-2)(\l'+3)}{2}}\begin{pmatrix}
\l&L&\l'\\
2&1&-3\end{pmatrix}+\sqrt{\frac{(\l'+2)(\l'-1)}{2}}\begin{pmatrix}
\l&L&\l'\\
2&-1&-1\end{pmatrix}\big]\nonumber,
\\
{}_{-2}I^{\phi}_{\l\l'LM} &=-\mathcal{W}_{\l L\l'}\sqrt{\frac{\Pi_{\l\l'L}L(L+1)}{8\pi}}
\big[\sqrt{\frac{(\l'-2)(\l'+3)}{2}}\begin{pmatrix}
\l&L&\l'\\
2&1&-3\end{pmatrix}+\sqrt{\frac{(\l'+2)(\l'-1)}{2}}\begin{pmatrix}
\l&L&\l'\\
2&-1&-1\end{pmatrix}\big]\nonumber,
\end{align}
and
\begin{align}
{}_{+2}I^{\Omega}_{\l\l'LM}&=i(-1)^{\l+L+\l'}\mathcal{W}_{\l L\l'}\sqrt{\frac{\Pi_{\l\l'L}L(L+1)}{8\pi}}
\big[\sqrt{\frac{(\l'-2)(\l'+3)}{2}}\begin{pmatrix}
\l&L&\l'\\
2&1&-3\end{pmatrix}-\sqrt{\frac{(\l'+2)(\l'-1)}{2}}\begin{pmatrix}
\l&L&\l'\\
2&-1&-1\end{pmatrix}\big]\nonumber,
\\
{}_{-2}I^{\Omega}_{\l\l'LM}&=-i\mathcal{W}_{\l L\l'}\sqrt{\frac{\Pi_{\l\l'L}L(L+1)}{8\pi}}
\big[\sqrt{\frac{(\l'-2)(\l'+3)}{2}}\begin{pmatrix}
\l&L&\l'\\
2&1&-3\end{pmatrix}\nonumber-\sqrt{\frac{(\l'+2)(\l'-1)}{2}}\begin{pmatrix}
\l&L&\l'\\
2&-1&-1\end{pmatrix}\big]\nonumber.
\end{align}
Finally, we define the coupling kernels as $H^{\psi}_{\l\l'L} = \sqrt{\frac{\Pi_{\l\l'L}}{4\pi}}\begin{pmatrix}
\l&L&\l'\\
2&0&-2\end{pmatrix}$ and $G^{\phi / \Omega}_{\l\l'L} = \sqrt{\frac{\Pi_{\l\l'L}L(L+1)}{8\pi}}\big[\sqrt{\frac{(\l'+2)(\l'-1)}{2}}
    \begin{pmatrix}
    \l&L&\l'\\
    2&-1&-1\end{pmatrix}
\pm \sqrt{\frac{(\l'-2)(\l'+3)}{2}}
\begin{pmatrix}
\l&L&\l'\\
2&1&-3\end{pmatrix}\big]$. The simplified expressions for each integral kernel can be conveniently derived, which can be written as ${}_{+2}I^{\psi}_{\l\l'LM} = (-1)^{\l+L+\l'}\mathcal{W}_{\l L\l'}H^{\psi}_{\l\l'L}$ and ${}_{-2}I^{\psi}_{\l\l'LM} = \mathcal{W}_{\l L\l'}H^{\psi}_{\l\l'L}$,  ${}_{+2}I^{\phi}_{\l\l'LM} = -(-1)^{\l+L+\l'}\mathcal{W}_{\l L\l'}G^{\phi}_{\l\l'L}$ and ${}_{-2}I^{\phi}_{\l\l'LM} = -\mathcal{W}_{\l L\l'}G^{\phi}_{\l\l'L}$, ${}_{+2}I^{\Omega}_{\l\l'LM} = -i(-1)^{\l+L+\l'}\mathcal{W}_{\l L\l'}G^{\Omega}_{\l\l'L}$ and  ${}_{-2}I^{\Omega}_{\l\l'LM} = i\mathcal{W}_{\l L\l'}G^{\Omega}_{\l\l'L}$.

\subsection{Perturbed E- and B-modes}
In the previous section, we obtained all the integral kernels. Therefore, in this section, we can compute the explicit forms of the E- and B-modes perturbed by chiral GWs. Considering second order perturbations, at leading order and second order, using Eq.~\eqref{E0} and Eq.~\eqref{E2}, we directly obtain $\E^{(0)}_{\l m}=E_{\l m}$, $\B^{(0)}_{\l m}=B_{\l m}$ and $\E^{(2)}_{\l m}=-\frac{1}{2}\left[(\l(\l+1)-4)R+4S\right]E_{\l m}$, $\B_{\l m}^{(2)}=-\frac{1}{2}\left[(\l(\l+1)-4)R+4S\right]B_{\l m}$. For the first order, using Eq.~\eqref{E1} and Eq.~\eqref{B1}, we can see that it is related to a linear combination of the three integral kernels. Using the parity operators $P^{+}_{\l\l'L}=[1+(-1)^{\l+\l'+L}]/2$ and $P^{-}_{\l\l'L}=[1-(-1)^{\l+\l'+L}]/2$, we obtain the respective combination forms as follows:
\begin{align}
    {}_{+2}I^{\psi}_{\l\l'LM} +{}_{-2}I^{\psi}_{\l\l'LM} &= 2\mathcal{W}_{\l L\l'}H^{\psi}_{\l\l'L}P^{+}_{\l\l'L}, \qquad {}_{+2}I^{\psi}_{\l\l'LM} -{}_{-2}I^{\psi}_{\l\l'LM} = -2\mathcal{W}_{\l L\l'}H^{\psi}_{\l\l'L}P^{-}_{\l\l'L},\nonumber
\\
    {}_{+2}I^{\phi}_{\l\l'LM} +{}_{-2}I^{\phi}_{\l\l'LM} &= -2\mathcal{W}_{\l L\l'}G^{\phi}_{\l\l'L}P^{+}_{\l\l'L}, \qquad {}_{+2}I^{\phi}_{\l\l'LM} -{}_{-2}I^{\phi}_{\l\l'LM} = 2\mathcal{W}_{\l L\l'}G^{\phi}_{\l\l'L}P^{-}_{\l\l'L},\nonumber
\\
    {}_{+2}I^{\Omega}_{\l\l'LM} +{}_{-2}I^{\Omega}_{\l\l'LM} &= 2i\mathcal{W}_{\l L\l'}G^{\Omega}_{\l\l'L}P^{-}_{\l\l'L}, \qquad {}_{+2}I^{\Omega}_{\l\l'LM} -{}_{-2}I^{\Omega}_{\l\l'LM} = -2i\mathcal{W}_{\l L\l'}G^{\Omega}_{\l\l'L}P^{+}_{\l\l'L}.\nonumber
\end{align}

Substituting the above combinations into Eq.~\eqref{E1} and Eq.~\eqref{B1}, we obtain the first-order perturbed E- and B-modes induced by chiral GWs, which can be written as
\begin{align}
\E^{(1)}_{\l m}&=\sum_{LM\l'm'} -\mathcal{W}_{\l L\l'}\left\{\phi_{LM}G^{\phi}_{\l\l'L}\left[E_{\l'm'}P^{+}_{\l\l'L}-iB_{\l'm'}P^{-}_{\l\l'L}\right]+i\left[\Omega_{LM}G^{\Omega}_{\l\l'L}-2\psi_{LM}H^{\psi}_{\l\l'L}\right]\left[E_{\l'm'}P^{-}_{\l\l'L}-iB_{\l'm'}P^{+}_{\l\l'L}\right]\right\},\label{deltaE}
\\
\B^{(1)}_{\l m}&=\sum_{LM\l'm'} -\mathcal{W}_{\l L\l'}\left\{\phi_{LM}G^{\phi}_{\l\l'L}\left[iE_{\l'm'}P^{-}_{\l\l'L}+B_{\l'm'}P^{+}_{\l\l'L}\right]+i\left[\Omega_{LM}G^{\Omega}_{\l\l'L}-2\psi_{LM}H^{\psi}_{\l\l'L}\right]\left[iE_{\l'm'}P^{+}_{\l\l'L}+B_{\l'm'}P^{-}_{\l\l'L}\right]\right\}.\label{deltaB}
\end{align}

\section{Correlations in CMB}
For the weak lensing effect induced by chiral GW, we denote the unperturbed CMB photon polarization modes as $E_{\l m}$ and $B_{\l m}$, and the perturbed modes as $\E_{\l m}$ and $\B_{\l m}$, which we have $\E_{\l m} = \E^{(0)}_{\l m} +  \E^{(1)}_{\l m}+ \E^{(2)}_{\l m}+\cdots$ and $\B_{\l m} = \B^{(0)}_{\l m} +  \B^{(1)}_{\l m}+ \B^{(2)}_{\l m}+\cdots$. Our goal is to obtain the EB correlation induced by chiral GWs. We consider the correlation up to second order, which can be written as $\langle \E^{*}_{\l m}\B_{\l m}\rangle = \langle \E^{(0)*}_{\l m}\B^{(0)}_{\l m}\rangle  +\langle  \E^{(1)*}_{\l m}\B^{(0)}_{\l m}\rangle +\langle \E^{(0)*}_{\l m} \B^{(1)}_{\l m}\rangle+\langle  \E^{*(1)}_{\l m} \B^{(1)}_{\l m}\rangle+\langle  \E^{(2)*}_{\l m}B^{(0)}_{\l m}\rangle +\langle \E^{(0)*}_{\l m} \B^{(2)}_{\l m}\rangle+\cdots.$

We assume that the EB correlation of CMB photons in the absence of chiral gravitational wave perturbations vanishes. Thus, the zeroth-order term  $\langle \E^{(0)*}_{\l m}\B^{(0)}_{\l m}\rangle=\langle E^{*}_{\l m}B^{}_{\l m}\rangle$ and the term $\langle  \E^{(2)*}_{\l m}B^{(0)}_{\l m}\rangle=\langle  \E^{(0)*}_{\l m}B^{(2)}_{\l m}\rangle=-\frac{1}{2}\left[(\l(\l+1)-4)R+4S\right]\langle E^{*}_{\l m}B_{\l m}\rangle$ these second order contributions vanished because $\langle E^{*}_{\l m}B_{\l m}\rangle=0$. For the first order term $\langle  \E^{(1)*}_{\l m}\B^{(0)}_{\l m}\rangle +\langle \E^{(0)*}_{\l m} \B^{(1)}_{\l m}\rangle$, in an isotropic background, condition $\langle\phi\rangle=\langle\Omega\rangle=\langle\psi\rangle=0$ is satisfied, and thus it naturally vanishes. In summary, up to the second order, the EB correlation induced by chiral GWs arises entirely from $\langle\E^{*(1)}_{\l m} \B^{(1)}_{\l m}\rangle$, which can be written as
\begin{align}
    \langle\E^{*(1)}_{\l m} \B^{(1)}_{\l m}\rangle &= -\frac{1}{2\l+1}\sum_{\l' L}\left[\langle\phi^{*}_{LM}\Omega_{LM}\rangle G^{\Omega}_{\l\l'L}G^{\phi}_{\l\l'L}-2\langle\phi^{*}_{LM}\psi_{LM}\rangle H^{\psi}_{\l\l'L}G^{\phi}_{\l\l'L}\right]\left[C^{EE}_{\l'}(P^{+}_{\l\l'L}-P^{-}_{\l\l'L})+C^{BB}_{\l'}(P^{-}_{\l\l'L}-P^{+}_{\l\l'L})\right],\nonumber
    \\
    &=-\frac{1}{2\l+1}\sum_{\l'L}(-1)^{\l+\l'+L}\left[C^{\Omega\phi}_{L} G^{\Omega}_{\l\l'L}G^{\phi}_{\l\l'L}-2C^{\phi\psi}_{L} H^{\psi}_{\l\l'L}G^{\phi}_{\l\l'L}\right]\left[C^{EE}_{\l'}-C^{BB}_{\l'}\right].
\end{align}
Finally, the resulting EB correlation of the CMB induced by chiral GWs is given by
\begin{align}
    \delta C^{EB}_{\l} = -\frac{1}{2\l+1}\sum_{\l'L}(-1)^{\l+\l'+L}\left[C^{\Omega\phi}_{L} G^{\Omega}_{\l\l'L}G^{\phi}_{\l\l'L}-2C^{\phi\psi}_{L} H^{\psi}_{\l\l'L}G^{\phi}_{\l\l'L}\right]\left[C^{EE}_{\l'}-C^{BB}_{\l'}\right].
\end{align}

Using the above formulas, we can calculate the EE and BB correlations, expanded to second order, the EE correlation is given by: $\langle \E^{*}_{\l m}\E_{\l m}\rangle = \langle \E^{(0)*}_{\l m}\E^{(0)}_{\l m}\rangle  +\langle  \E^{(1)*}_{\l m}\E^{(0)}_{\l m}\rangle +\langle \E^{(0)*}_{\l m} \E^{(1)}_{\l m}\rangle+\langle  \E^{*(1)}_{\l m} \E^{(1)}_{\l m}\rangle+\langle  \E^{(2)*}_{\l m}E^{(0)}_{\l m}\rangle +\langle \E^{(0)*}_{\l m} \E^{(2)}_{\l m}\rangle+\cdots.$ Likewise, for the first order term $\langle  \E^{(1)*}_{\l m}\E^{(0)}_{\l m}\rangle +\langle \E^{(0)*}_{\l m} \E^{(1)}_{\l m}\rangle$  also vanishes in isotropic background. The zeroth-order contribution is equivalent to the primordial EE spectrum, whereas the second-order correction introduces $\langle  \E^{(2)*}_{\l m}E^{(0)}_{\l m}\rangle +\langle \E^{(0)*}_{\l m} \E^{(2)}_{\l m}\rangle=-[(l(l+1)-4)R+4S]\langle E_{\l m}^{*} E_{\l m}\rangle$. Finally, using Eq.~\eqref{deltaE}, we naturally obtain
\begin{align}
\begin{split}
    \delta C^{EE}_{\l} = -[(\l&(\l+1)-4)R+4S]C^{EE}_{\l}+\frac{1}{2\l+1}\sum_{\l'L}\{C^{\phi\phi}_{L}(G^{\phi}_{\l\l'L})^{2}(C^{EE}_{\l'}P^{+}_{\l\l'L}+C^{BB}_{\l'}P^{-}_{\l\l'L})
    \\
    &+\left[C^{\Omega\Omega}_{L}(G^{\Omega}_{\l\l'L})^{2}-4C^{\Omega\psi}_{L}G^{\Omega}_{\l\l'L}H^{\psi}_{\l\l'L}+4C^{\psi\psi}_{L}(H^{\psi}_{\l\l'L})^{2}\right](C^{EE}_{\l'}P^{-}_{\l\l'L}+C^{BB}_{l'}P^{+}_{\l\l'L})\}.
\end{split}
\end{align}
In the end, using the same method, we can obtain the form of the chiral gravitational wave perturbation to the BB spectrum, which can be expressed as
\begin{align}
    \begin{split}
    \delta C^{BB}_{\l} = -[(\l&(\l+1)-4)R+4S]C^{BB}_{\l}+\frac{1}{2\l+1}\sum_{\l'L}\{C^{\phi\phi}_{L}(G^{\phi}_{\l\l'L})^{2}(C^{EE}_{\l'}P^{-}_{\l\l'L}+C^{BB}_{l'}P^{+}_{\l\l'L})
    \\
    &+\left[C^{\Omega\Omega}_{L}(G^{\Omega}_{\l\l'L})^{2}-4C^{\Omega\psi}_{L}G^{\Omega}_{\l\l'L}H^{\psi}_{\l\l'L}+4C^{\psi\psi}_{L}(H^{\psi}_{\l\l'L})^{2}\right](C^{EE}_{\l'}P^{+}_{\l\l'L}+C^{BB}_{\l'}P^{-}_{\l\l'L})\}.
\end{split}
\end{align}

\subsection{Temperature perturbation}
In this subsection, we investigate the temperature perturbation induced by the lensing effect of chiral GWs. It can be expressed as $\T(\n) \Longrightarrow \T(\n+\delta n)=\widetilde{\T}(\n) = \T(\n)+\delta n ^{i} \nabla_{i} \T (\n) +\frac{1}{2}\delta n^{i}\delta n^{j}\nabla_{i}\nabla_{j}\T(\n)+\cdots.$ Expanding this in spherical harmonics, we obtain $\widetilde{\T} = \sum_{\l m}\widetilde{T}_{\l m}Y_{\l m}$, which means $\widetilde{\T}_{\l m} = \int \dd \n Y^{*}_{\l m} \widetilde\T=\int \dd \n Y^{*}_{\l m}\left[\T(\n)+\delta n ^{i} \nabla_{i} \T (\n) +\frac{1}{2}\delta n^{i}\delta n^{j}\nabla_{i}\nabla_{j}\T(\n)\right]$. Substituting $\T = \sum_{\l'm'}\T_{\l'm'}Y_{\l'm'}$ into the above expression, we obtain the leading order and second order perturbations as follows, $\widetilde{\T}^{(0)}_{\l m}=\sum_{\l'm'}\T_{\l'm'}\int \dd \n Y^{*}_{\l m}Y_{\l'm'}=\T_{\l m}$ and $\widetilde{\T}^{(2)}_{\l m}=\frac{1}{2}R\sum_{\l'm'}\T_{\l'm'}\int \dd \n Y^{*}_{\l m}\nabla^{2}Y_{\l'm'}=-\frac{R}{2}(\l(\l+1))\T_{\l m}$. Similarly, the first-order perturbation is given by $\widetilde{\T}^{(1)}_{\l m}=\int \dd \n Y^{*}_{\l m} \left\{\left( \nabla^{i}\phi - \epsilon^{ijk}n_{j}\nabla_{k}\Omega \right)\nabla_{i}\T\right\}$. Simultaneously expanding $\phi$, $\Omega$ and $\T$ , we obtain $\widetilde\T_{\l m}^{(1)}=\sum\left[\phi_{LM}\T_{\l'm'}I^{\T\phi}_{\l \l'LM}-\Omega_{LM}\T_{\l'm'}I^{\T\Omega}_{\l\l'LM}\right]$, where
\begin{align}
    I^{\T\phi}_{\l\l'LM}=\int \dd\n \left[Y^*_{\l m}\nabla^{i}Y_{LM}\nabla_iY_{\l'm'}\right],\qquad
    I^{\T\Omega}_{\l\l'LM}=\int \dd\n \left[Y^*_{\l m}\epsilon^{ijk}n_j\nabla_kY_{LM}\nabla_iY_{\l'm'}\right].\nonumber
\end{align}
The calculation of these two integral kernels follows the same procedure as in Appendix~\ref{Integral Kernels}. Utilizing the Gaunt integral and the Newman-Penrose formalism, we obtain:
\begin{align}
    I^{\T\phi}_{\l\l'LM}&=  \frac{1}{2}\int \dd\n Y^*_{\l m}\left(m^{i}\bar\eth+\bar m^{i}\eth\right)Y_{LM}\left(m_{i}\bar\eth+\bar m_{i}\eth\right)Y_{\l'm'}=-\mathcal{W}_{\l L\l'}\sqrt{L(L+1)\l'(\l'+1)}\sqrt{\frac{\Pi_{\l\l'L}}{4\pi}}\begin{pmatrix}
\l&L&\l'\\
0&-1&1\end{pmatrix}P^{+}_{\l\l'L},
\\
I^{\T\Omega}_{\l\l'LM} &= -\frac{i}{2}\int \dd\n Y^*_{\l m}\left(m^{i}\bar\eth-\bar m^{i}\eth\right)Y_{LM}\left(m_{i}\bar\eth+\bar m_{i}\eth\right)Y_{\l'm'} = i\mathcal{W}_{\l L\l'}\sqrt{L(L+1)\l'(\l'+1)}\sqrt{\frac{\Pi_{\l\l'L}}{4\pi}}\begin{pmatrix}
\l&L&\l'\\
0&-1&1\end{pmatrix}P^{-}_{\l\l'L}.
\end{align}
Here, we defined inner coupling kernel $K_{\l\l'L}=-\sqrt{L(L+1)\l'(\l'+1)}\sqrt{\frac{\Pi_{\l\l'L}}{4\pi}}\begin{pmatrix}
\l&L&\l'\\
0&-1&1\end{pmatrix}$, and the temperature perturbation induced by chiral GWs can be written as
\begin{align}
    \widetilde{\T}_{\l m} = \T_{\l m}-\frac{R}{2}[\l(\l+1)]\T_{\l m} +\sum\mathcal{W}_{\l L\l'}\left[\phi_{LM}K_{\l\l'L}\T_{\l'm'}P^{+}_{\l\l'L}+i\Omega_{LM}K_{\l\l'L}\T_{\l'm'}P^{-}_{\l\l'L}\right].
\end{align}
Having obtained the explicit form of the temperature perturbation induced by chiral GWs, we can calculate the associated correlations. We compute up to the second-order correlation. Similarly, terms involving $\langle\phi\rangle,\langle\Omega\rangle,\langle\psi\rangle$ naturally vanish in the isotropic background, and the temperature auto correlation perturbation can be expressed as
\begin{align}
    \delta C^{\T\T}_{\l} = -\l(\l+1)RC^{\T\T}_{\l} + \frac{1}{2\l+1} \sum_{\l'L}\left\{C^{\phi\phi}_{L}(K_{\l\l'L})^{2}C^{\T\T}_{l'}P^{+}_{\l\l'L}+C^{\Omega\Omega}_{L}(K_{\l\l'L})^{2}C^{\T\T}_{\l'}P^{-}_{\l\l'L}\right\}.
\end{align}

At this point, we can introduce another TB correlation associated with parity violation. Assuming the primordial TB correlation vanishes and the background is isotropic, the TB correlation induced by chiral GWs can be written as  $\langle \widetilde{\T}^{*}_{\l m}\B_{\l m}\rangle =  \langle  \widetilde{\T}^{*(1)}_{\l m} \B^{(1)}_{\l m}\rangle$,
\begin{align}
    \delta C^{\T B}_{\l} = -\frac{1}{2\l+1} \sum_{\l'L}\left\{C^{\phi\Omega}_{L}G^{\phi}_{\l\l'L}K_{\l\l'L}C^{\T E}_{\l'}P^{-}_{\l\l'L}-\left[C^{\phi\Omega}_{L}G^{\Omega}_{\l\l'L}K_{\l\l'L}-2C^{\phi\psi}_{L}H^{\psi}_{\l\l'L}K_{\l\l'L}\right]C^{\T E}_{\l'}P^{+}_{\l\l'L}\right\}.
\end{align}
Finally, for the even-parity TE correlation, assuming the unperturbed TE correlation is non-zero, the perturbed TE correlation can be expressed as $\langle \widetilde{\T}^{*}_{\l m}\E_{\l m}\rangle = \langle\widetilde{\T}^{*(0)}_{\l m} \E^{(0)}_{\l m}\rangle +  \langle\widetilde{\T}^{*(1)}_{\l m} \E^{(1)}_{\l m}\rangle +\langle\widetilde{\T}^{*(0)}_{\l m} \E^{(2)}_{\l m}\rangle + \langle\widetilde{\T}^{*(2)}_{\l m} \E^{(0)}_{\l m}\rangle$,
\begin{align}
    \delta C^{\T E}_{\l} = -[\l&(\l+1)-2]RC^{\T E}_{\l} - 2SC_{\l}^{TE}\nonumber
    \\
    &-\frac{1}{2\l+1}\sum_{\l'L}\{C^{\phi\phi}_{L}G^{\phi}_{\l\l'L}K_{\l\l'L}C^{\T E}_{\l'}P^{+}_{\l\l'L}+\left[C^{\Omega\Omega}_{L}G^{\Omega}_{\l\l'L}K_{\l\l'L}-2C^{\Omega\psi}_{L}H^{\psi}_{\l\l'L}K_{\l\l'L}\right]C^{\T E}_{\l'}P^{-}_{\l\l'L}\}.
\end{align}

\subsection{$C^{\Omega\phi}_{L}$ and $C^{\phi\psi}_{L}$ from chiral GWB}
The formula for $C^{\Omega\phi}_{L}$ can be expressed in the following integral form~\cite{Inomata:2024ald,book2012odd} 
\begin{align}\label{C_phiomega}
    C^{\phi\Omega}_{L} = \int^{k_{max}}_{k_{min}} \frac{k^{2}\dd k}{2\pi^{2}}\left[P_{\text{L}}(k)-P_{\text{R}}(k)\right]F^{\phi}_{L}(k)F^{\Omega}_{L}(k),
\end{align}
where, the integral kernel with respect to $\phi$ and $\Omega$ can be written as
\begin{align}
    F^{\Omega}_{L}(k) &= N_{L} \int^{k\eta_0}_{k\eta_{r}} \dd \omega T(\omega)\frac{j_{L}(k\eta_{0}-\omega)}{(k\eta_{0}-\omega)^{2}},
    \\
    F_{L}^{\phi}(k) &= -N_{L}\int^{k\eta_0}_{k\eta_{r}}\dd\omega T(\omega)\frac{1}{k(\eta_{0}-\eta_{r})}\left\{(k\eta_{0}-\omega)\left[\frac{\partial}{\partial\omega}+\frac{1}{2}(\omega-k\eta_{r})(1+\frac{\partial^{2}}{\partial\omega^{2}})\right]-3-2(\omega-k\eta_{r})\frac{\partial}{\partial\omega}\right\}\frac{j_{L}(k\eta_{0}-\omega)}{(k\eta_{0}-\omega)^{2}}.
\end{align}
Here, $\omega=k\eta$, $N_{L}=\frac{1}{L(L+1)}\sqrt{2\pi\frac{(L+2)!}{(L-2)!}}$ and $T(\omega)=3j_{1}(\omega)/\omega$.  We set $k\eta_{0}-\omega=x$ and $\omega-k\eta_{r}=y$, the $\frac{\partial}{\partial\omega}=\frac{\partial}{\partial x}\frac{\partial x}{\partial \omega}=-\frac{\partial}{\partial x}$, therefore, we have
\begin{align}
    F^{\phi}_{L} &=-N_{L}\int^{k\eta_0}_{k\eta_{r}}\dd\omega \frac{T(\omega)}{k(\eta_{0}-\eta_{r})}
    \left\{x[-\frac{\partial}{\partial x}+\frac{y}{2}(1+\frac{\partial^{2}}{\partial x^{2}})]-3+2y\frac{\partial}{\partial x}\right\}\frac{j_{L}(x)}{x^{2}}\nonumber
     \\
    & =-N_{L}\int^{k\eta_0}_{k\eta_{r}}\dd\omega \frac{T(\omega)}{k(\eta_{0}-\eta_{r})}\left\{\frac{k(\eta_{0}-\eta_{r})j_{L+1}(k\eta_{0}-\omega)}{(k\eta_{0}-\omega)^{2}} -\frac{(L+1)j_{L}(k\eta_{0}-\omega)}{(k\eta_{0}-\omega)^{2}}   + \frac{(\omega-k\eta_{r})}{2}\frac{(L+1)(L-2)j_{L}(k\eta_{0}-\omega)}{(k\eta_{0}-\omega)^{3}}\right\}.
\end{align}
On the other hand, $\psi$ and $\Omega$ satisfy the relation $\psi = -\frac{1}{2}\nabla^{2}\Omega$, which reflects the general effect of metric perturbations on a bundle of photons propagating along geodesics. Using this condition, we find that their expansions must satisfy $\sum_{LM}\psi_{LM}Y_{LM}=-\frac{1}{2}\sum_{LM}\Omega_{LM}\nabla^{2}Y_{LM}$, which means $\psi_{LM}=\frac{L(L+1)}{2}\Omega_{LM}$. Furthermore, the relation between $\psi$ and $\phi$ can be expressed as $C^{\phi\psi}_{L} = \frac{L(L+1)}{2}C^{\phi\Omega}_{L}$. Finally, by defining a dimensionless chiral spectrum $\mathcal{P}_{V}(k)=\frac{k^{3}}{2\pi^{2}}\left[P_{\mathcal{L}}(k)-P_{\mathcal{R}}(k)\right]$ to characterize the degree of parity violation in chiral GWs, $C^{\phi\Omega}_{L}$ and $C^{\phi\psi}_{L}$ can be respectively written as $C^{\phi\Omega}_{L}=\int \dd \text{ln}k \ \mathcal{P}_{V}(k)F^{\phi}_{L}(k)F^{\Omega}_{L}(k)$ and $C^{\phi\psi}_{L}=\frac{L(L+1)}{2}\int \dd \text{ln}k \ \mathcal{P}_{V}(k)F^{\phi}_{L}(k)F^{\Omega}_{L}(k)$. By substituting $C^{\phi\Omega}_{L}$ and $C^{\phi\psi}_{L}$ calculated by Eq.~\eqref{C_phiomega} into Eq.\eqref{C_EB}, we can obtain $C^{EB}_{\l}=\int\dd\text{ln}k\mathcal{P}_{V}(k)\mathcal{K}^{EB}_{\l}(k)$ and the integral kernel for the EB correlation
\begin{align}
    \mathcal{K}^{EB}_{\l}(k) = -\frac{1}{2\l+1}\sum_{\l'L}(-1)^{\l+\l'+L}\left[G^{\Omega}_{\l\l'L}G^{\phi}_{\l\l'L}- L(L+1)H^{\psi}_{\l\l'L}G^{\phi}_{\l\l'L}\right]\left[F^{\phi}_{L}(k)F^{\Omega}_{L}(k)\right] \left[C^{EE}_{\l'}-C^{BB}_{\l'}\right],
\end{align}
which governs the shape of the EB correlation curve induced by chiral GWs. Moreover, the integral kernel for the BB  also can similarly be written as
\begin{align}
\begin{split}
    \mathcal{K}^{BB}_{\l}(k) =& -[(\l(\l+1)-4)r+4s]C^{BB}_{\l}+\frac{1}{2\l+1}\sum_{\l'L}\{(G^{\phi}_{\l\l'L})^{2}(F^{\phi}_{L}(k))^{2}(C^{EE}_{\l'}P^{-}_{\l\l'L}+C^{BB}_{\l'}P^{+}_{\l\l'L})
    \\   &+\left[G^{\Omega}_{\l\l'L}-L(L+1)H^{\psi}_{\l\l'L}\right]^{2}(F^{\Omega}_{L}(k))^{2}(C^{EE}_{\l'}P^{+}_{\l\l'L}+C^{BB}_{\l'}P^{-}_{\l\l'L})\},
\end{split}
\end{align}
where $r = \sum L(L+1)(2L+1)\left[(F^{\phi}_{L})^{2}+(F^{\Omega}_{L})^{2}\right]/8\pi$ and $s = \sum (2L+1)L^{2}(L+1)^{2}(F^{\Omega}_{L})^{2}/16\pi$. It also governs the shape of the BB correlation. The wavenumber $k$ determines the integration interval and thus affects only the amplitude of the curve,  the resulting EB angular power spectra have similar shapes over the plotted multipole range.

\twocolumngrid
\let\addcontentsline\savedaddcontentsline
\bibliography{myrefs.bib}	
\end{document}